\documentclass[
    ,final            
  ]
  {aipproc}
\usepackage{amsmath,amssymb}

\layoutstyle{6x9}

\def\jour#1#2#3#4{{#1} {\bf#2} (19#3) #4}
\def\jou2#1#2#3#4{{#1} {\bf#2} (20#3) #4}
\def\AP{Acta Phys. Pol. {B}}

\def\EPJ{Eur. Phys. J. {C}}

\def\IAN{Izv. Akad. Nauk: Ser. Fiz.}  

\def\JP{J. Phys. {G}}

\def\NC{Nuovo Cim. {A}}

\def\NPA{Nucl. Phys. {A}}
\def\NP{Nucl. Phys. {B}}

\def\PL{Phys. Lett.  {B}}
\def\PRC{Phys. Rev. {C}}
\def\PRD{Phys. Rev. {D}}
\def\PRL{Phys. Rev. Lett.}
\def\PRp{Phys. Rep.}

\def\ZP{Z. Phys.  {C}}
\def\ep{$\rm{e}^+\rm{e}^-$}
\def\nopar{\noindent} 

\def\Ecmn{\sqrt{s_{\rm NN}}}
\def\Ecmp{\sqrt{s_{\rm pp}}}
\def\Ecme{\sqrt{s_{\rm ee}}}
\def\ep{$\rm{e}^+\rm{e}^-$}
\def\app{${\bar {\rm p}}{\rm p}$}
\def\timesplus{\times \hspace{-.275cm}+}
\def\triangleinsquare{\square\hspace*{-.275cm} \scriptstyle \triangle}
\def\blktriangdowninsqr{ \square\hspace*{-.265cm} \blacktriangledown}
\def\blktrianginsqr{\square\hspace*{-.27cm} \blacktriangle}
\def\timesplussquared{{ \scriptstyle \boxtimes} \hspace{-.24cm}+}
\def\vs{\vspace*}
 \def\bi{\bibitem}

\def\col{Collab.}
\def\ea{{\sl et al.}}
\def\eg{{\sl e.g.}}
\def\vrs{{\sl vs.}}

\def\ct{\cite}

\def\gtsim{\gtrsim}
\begin{document}

\title{Multihadron production features\\ in 
 different reactions}


\author{Edward K.G. Sarkisyan}{
    address={EP Division, Department of Physics, CERN, CH-1211 Geneva 
23, Switzerland}
 ,altaddress={Department of Physics, the University of Manchester,
Manchester M13
9PL, UK}
}

\author{Alexander S. Sakharov}{
 address={TH Division, Department of Physics, CERN, CH-1211 Geneva 23, 
Switzerland},
altaddress={Swiss Institute of Technology, ETH-Z\"urich, 8093 
Z\"urich,
Switzerland
}
}

\begin{abstract}
 We consider multihadron production processes in different types of 
collisions in the framework of the picture based on   
dissipating energy of 
participants and their types.
 In particular, 
 the similarities of such bulk
observables like the charged particle  mean multiplicity and
the  pseudorapidity density at midrapidity measured
 in nucleus-nucleus, (anti)proton-proton and 
electron-positron 
interactions are
analysed.
 Within the description proposed a
good agreement
with the measurements   
in a wide range of  nuclear collision
energies from
AGS to RHIC is obtained. The predictions  up to the LHC energies are 
made
and compared to
experimental extrapolations.
\end{abstract}

\maketitle




 {\bf 1.} 
High 
densities and 
temperatures of nuclear matter reached at RHIC 
provide us with an exceptional  
opportunity to 
investigate the matter at 
 extreme conditions. 
 Bulk observables such as
 multiplicity and particle densities (spectra) being  
 sensitive to the dynamics of strong interactions, are of fundamental 
interest.
 Recent measurements at RHIC revealed striking evidences in the hadron 
production process including
 similarity
in  such  basic observables like the mean 
multiplicity and the midrapidity density measured
in complex  ultra-relativistic nucleus-nucleus (AA) collisions \vrs\ 
those 
obtained 
in relatively ``elementary'' 
\ep\ interactions
at the same centre-of-mass (c.m.) energy when  
 number of participants 
(``wounded''  nucleons \cite{woundN} 
in  AA
collisions) 
are taken into account
\ct{ph-rev,uni}. 
The observation is shown to be independent of the  
c.m. energy per nucleon $\Ecmn =$ 19.6 GeV to 200 GeV.
 Assuming similar mechanisms of hadron production in both types of 
interactions which then depends only on the amount of energy 
transformed into 
particles produced, one would expect the same value of the observables to 
be 
obtained in hadron-hadron collisions 
 at close c.m. energies. 
 However, this is not the case: comparing measurements in hadronic data 
 \ct{ua5-53900,cdf} to
the findings at RHIC, 
one obtains 
\ct{ph-rev,br-rev,phx-rev}
quite lower values in hadron-hadron collisions.
In the meantime, the RHIC dAu data 
at 
$\Ecmn=$ 200~GeV unambiguously point to the values of the mean 
multiplicity from \app\ data  
\ct{ph-rev}.
Moreover, recent CuCu RHIC data show no changes in the values of 
the  bulk 
variables compared to those from AuAu collisions 
when
properly normalised to the number of participants \ct{rachid,CuCu}.

 The observations made earlier \ct{ph-rev} and the recent ones \ct{CuCu} 
can be understood in the franework of a description  proposed recently by 
us \ct{my} and considered here. This description is  based on a picture 
when the whole
process of a collision is interpreted as the expansion and break-up
into particles of an initial state, in which the whole available energy is
assumed to be concentrated in a small Lorentz-contracted volume.
 There are no any restrictions 
due to 
the conservation of quantum numbers besides energy 
and
momentum constraints allowing therefore to link the amount of energy 
deposited in 
the collision zone and features of bulk variables in different reactions. 
 This description resembles the Landau
hydrodynamical approach to multiparticle production 
\cite{landau} which
 has been found to give good description  of 
the mean multiplicity
AA, pp,
\ep, $\nu$p data \ct{feinberg1,landau-exp} as well as  
 of pseudorapidity 
distributions at RHIC \ct{br-rev}. 

 As soon as a collision of two Lorentz-contracted
particles leads to the full thermalization of the system before 
extension, 
one
can assume that the production of secondaries is defined by the fraction
of participants energy deposited in the volume of the system at 
the collision
moment. This implies that there is a difference between results of 
collisions of structureless particles like electron and 
composite 
particles
like proton, the latter considered to be built of  
constituents. 
 Indeed, in composite particle collisions
not all the constituents deposit their energy when they form the
Lorentz-contracted volume of the thermalized initial state.
As a result, the leading particles 
\ct{leadp}, formed out 
of
those constituents which are not trapped in the interaction volume, carry
away a part of energy. 
Meantime, colliding structureless
particles are ultimately stopped as a whole in the initial state of
the thermalized collision zone depositing their total energy in the
Lorentz-contracted volume and this energy is 
wholly available for production of secondaries.

We consider a single nucleon as a superposition of three 
constituent
quarks due to the additive quark picture \ct{constq}. In this picture, 
most often only one quark from each nucleon 
contributes to the interaction with other
quarks being spectators.  Thus, the initial thermalized 
state 
is pumped in
only by the energy of the interacting single quark pair and, so,  
only 1/3 of the entire nucleon energy is available for
production of secondaries.
Therefore, 
one expects
that the resulting bulk variables like the multiplicity and rapidity
distributions should show identical features in \app\ collisions
at the c.m. energy $\Ecmp$ and in \ep\ interactions at the c.m. energy 
$\Ecme \simeq\Ecmp/3$.
 Note that for the mean multiplicity, a similar behaviour was found
in 
the beginning of LEP activity \ct{ee3pp}. 

In AA collisions, more than one quark
per nucleon interacts due to the large size of nucleus
and the long path of interactions inside the nucleus. 
 In central AA collisions, a contribution of constituent quarks 
rather than participating nucleons
seem to determine the properties of produced particle distributions 
\cite{voloshin}.
In 
headon collisions, the density of matter is 
almost saturated, so that all three constituent quarks 
from each nucleon may participate 
nearly simultaneously 
in collision 
depositing their 
energy coherently into the thermalized zone. 
Therefore, in the 
headon
AA
interactions at $\Ecmn$  the bulk variables are expected to have the values 
similar to those  
from
pp collisions at $\Ecmp \simeq 
3\, \Ecmn$.
This makes the most central
collisions of nuclei akin to \ep\ collisions 
at $\Ecme \simeq \Ecmn$ in sense
of the resulting bulk variables. 
 \smallskip

 \begin{figure}
\vspace*{-.1cm}
  \includegraphics[height=.54\textheight]{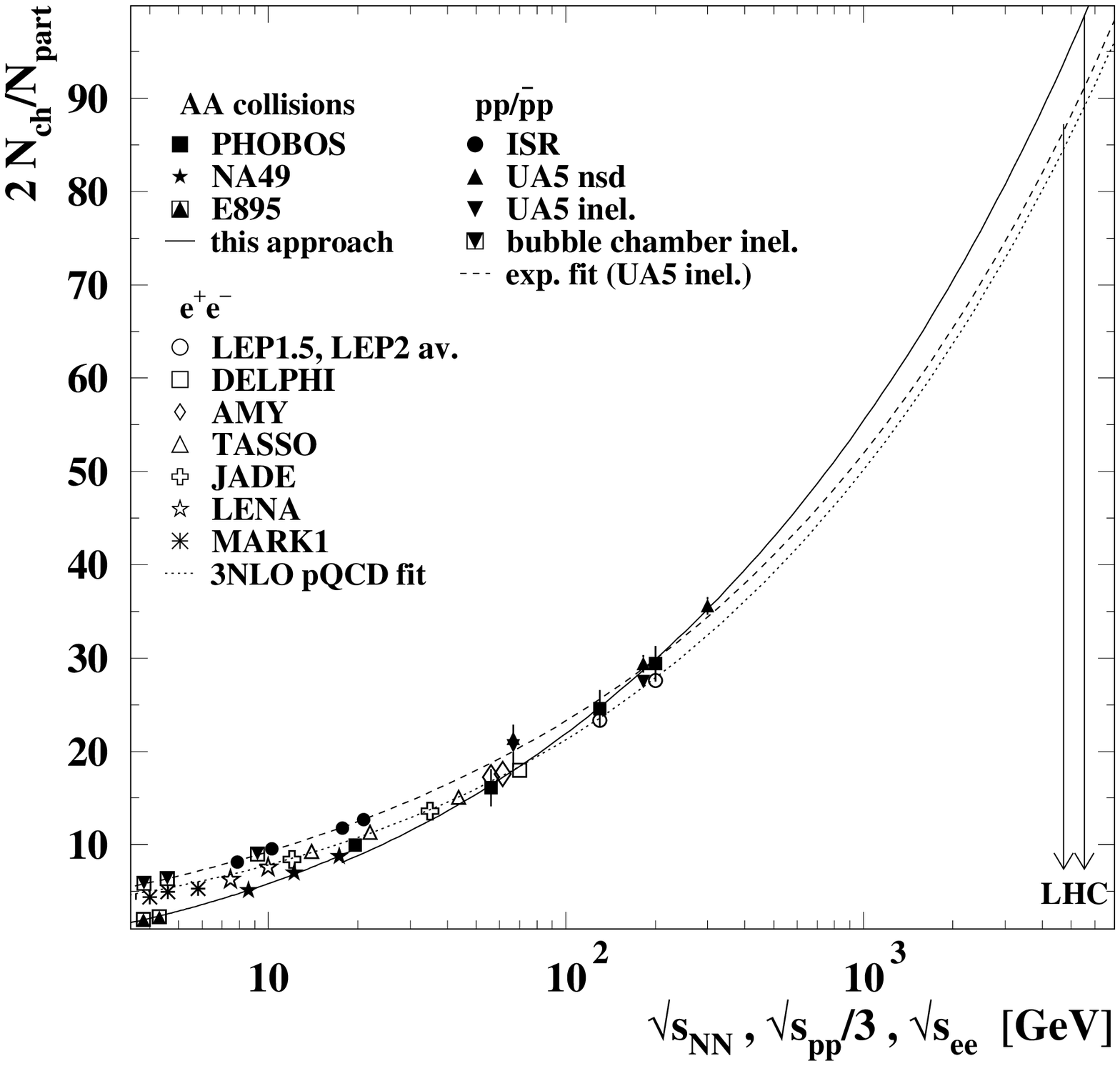}
 \vs{-.6cm}
\caption{
 The charged particle mean multiplicity $N_{\rm ch}$ per participant pair 
($N_{\rm 
part}$/2)
as a function of the c.m. energy. 
 The solid and combined symbols show the multiplicity values from: 
  most central 
heavy-ion (AA)
collisions  
\vrs\ c.m. energy per nucleon, $\Ecmn$,
  measured by PHOBOS 
\ct{ph-rev}
($ \scriptstyle 
\blacksquare$), 
NA49 \ct{na49-mult} 
($ \scriptstyle 
 \bigstar$), 
and  E895 \ct{agsmult} 
($\blktrianginsqr$)
(see also \ct{ph-rev});
 \app\ collisions, by  
 UA5  ($\blacktriangle$ for non-single diffractive, 
$\blacktriangledown$ for inelastic events) at 
$\Ecmp=$ 546 GeV
\ct{ua5-546} and $\Ecmp=$ 200 and 900 GeV \ct{ua5-zp}; 
 pp collisions (at lower $\Ecmp$) from CERN-ISR 
($\bullet$) \ct{isr-thome} and bubble chamber experiments 
\ct{bubblechamber,fnalmult} ($\blktriangdowninsqr$) (the latter 
compiled and analysed
in 
\ct{eddi}). 
 The inelastic UA5 data at $\Ecmp=$~200~GeV is due to the extrapolation 
in
 \ct{ph-rev}.
 The open symbols show the \ep\ measurements:  the 
high-energy LEP 
mean 
multiplicities ($ \scriptscriptstyle 
\bigcirc$) 
averaged  here from the
data
at 
 LEP1.5 $\Ecme =$~130~GeV  \ct{lep1.5o,lep1.5-2adl} and LEP2
$\Ecme
=$~200~GeV 
\ct{lep1.5-2adl,lep2o}, and the lower-energy data  by DELPHI 
\ct{delphi70} ($ \scriptstyle 
\square$), 
TASSO  ($ \scriptstyle 
 \triangle$), 
AMY   ($ \scriptstyle 
 \diamondsuit$), JADE  (+), LENA  
($\star$), and MARK1 
 ($\timesplus$) 
experiments. 
 (See 
refs. in \ct{opalcomp,pdg,biebel} 
for  \ep\ and pp/\app\ data).
 The solid line shows the calculations from Eq.~(\ref{prap0}) 
based on our approach and using the corresponding fits (see text).
 The dashed and dotted lines show the fit \ct{ua5-546} to the pp/\app\  
data
and 
the 3NLO  perturbative QCD  \ct{dremin} ALEPH fit \ct{lep1.5-2adl}
 to \ep\ data.
 The arrows show the LHC expectations.
 \vspace*{-.7cm}
 }
\label{fig:multshe}
\end{figure}

{\bf 2.} 
 According to our consideration, in Fig.~\ref{fig:multshe}, we compare  
  the c.m. energy dependence of  
the mean multiplicity  in AA 
and \ep\ interactions 
to 
that 
in 
 pp/\app\
collisions at $\Ecmn = \Ecme = 
\Ecmp/3$ 
from a few GeV 
to 200~GeV.
For $\Ecme> M_{Z^0}$, we give 
the 
multiplicities averaged  \ct{pdg} from  the recent    
LEP data at $\Ecme =$~130~GeV and 200~GeV:  
$23.35\pm 0.20 \pm 0.10$  and 
$27.62 \pm 0.11 \pm 0.16$. 
 Figure shows also the mean multiplicity fit to pp/\app\    
data \ct{ua5-546}  and the 3NLO pQCD \ct{dremin} ALEPH fit to \ep\ data 
\ct{lep1.5-2adl}.

From Fig.~\ref{fig:multshe} one sees that the pp/\app\ data 
  are very 
close to 
the \ep\ data at 
$\Ecme=\Ecmp/3$. 
This nearness  
decreases the 
already small deficit in the \ep\ data 
as the energy increases. 
 The deviation can be attributed to the inelasticity factor, or leading 
particle effect \cite{leadp} in pp/\app\ collisions, which is known to 
decrease 
with the 
c.m. energy. Then, at lower $\Ecmp$, some fraction of the energy of 
spectators contributes more into the formation of the initial state as 
the 
spectators pass by. This leads to the excess of the mean multiplicity 
in pp/\app\ data compared to the \ep\ data  
as it is seen in  Fig.~\ref{fig:multshe}.
 Comparing further 
the average 
 multiplicities from pp/\app\ collisions to
those from AA ones,  one finds that the 
data points  are amazingly close 
 to each other when the AA data are confronted the hadronic 
data 
at
$\Ecmp  =3\, \Ecmn$.
The inclusion of the tripling energy factor 
indeed allows to describe 
such a fundamental variable as the mean 
multiplicity  
{\it simultaneously} in 
\ep, pp/\app\ and central AA collisions for all energies. 
This shows that the  multiparticle production process in  headon 
AA collisions 
 is 
derived by the energy deposited in the Lorentz-contracted volume by 
a single pair of effectively 
structureless nucleons similar to that 
 in \ep\ annihilation and of quark-pair interactions in pp/\app\ 
collisions.   
 Note that an examination of  Fig.~\ref{fig:multshe} reveals 
that not  
a factor 1/2 
is needed to rescale  
$\Ecmp$  
to match the AA or \ep\ data as earlier was assumed   
for the mean multiplicity 
while
  recognised  to  unreasonably shift the \ep\ data
 on the 
pseudorapidity density at 
midrapidity when compared to the AA measurements
\ct{ph-rev}. 
This 
discrepancy  
 finds its explanation in our consideration, within which the 
data on $\it 
both$ the mean 
multiplicity and the midrapidity density ({\sl vide infra}) are 
self-consistently
 matched for different reactions.  
Let us gain recall a factor 1/3 obtained earlier in
\ct{ee3pp} for $\Ecmp$ for the  pp mean multiplicity data relative 
to 
those 
from \ep\
data, similar to our finding.

  Fig.~\ref{fig:multshe} shows  that 
the mean multiplicities in  different  reactions 
are close starting from the SPS $\Ecmn$ , and 
become particularly close  
at 
$\Ecmn \gtsim$~50~GeV.  However, at lower energies, 
the AA  data
are slightly 
below  the  \ep\ and hadronic data and the 
nuclear data
increase  faster with
energy than the pp and \ep\ 
data do.   On the 
other hand, as the c.m.
energy increases above a few tens GeV, the AA data start to overshoot
the \ep\ data and 
reach the mean multiplicity values from \app\ interactions. 
 From this one concludes on  two different energy regions of the
multiparticle production in AA reactions.
 The observations made can be understood in terms of the overlap zone and 
energy deposition by participants \ct{my}.
 Due to this,
one
would expect 
the differences to be more pronounced in midrapidity 
densities as discussed  below.
 \smallskip

\begin{figure}
 \vspace*{-.6cm}
  \includegraphics[height=.54\textheight]{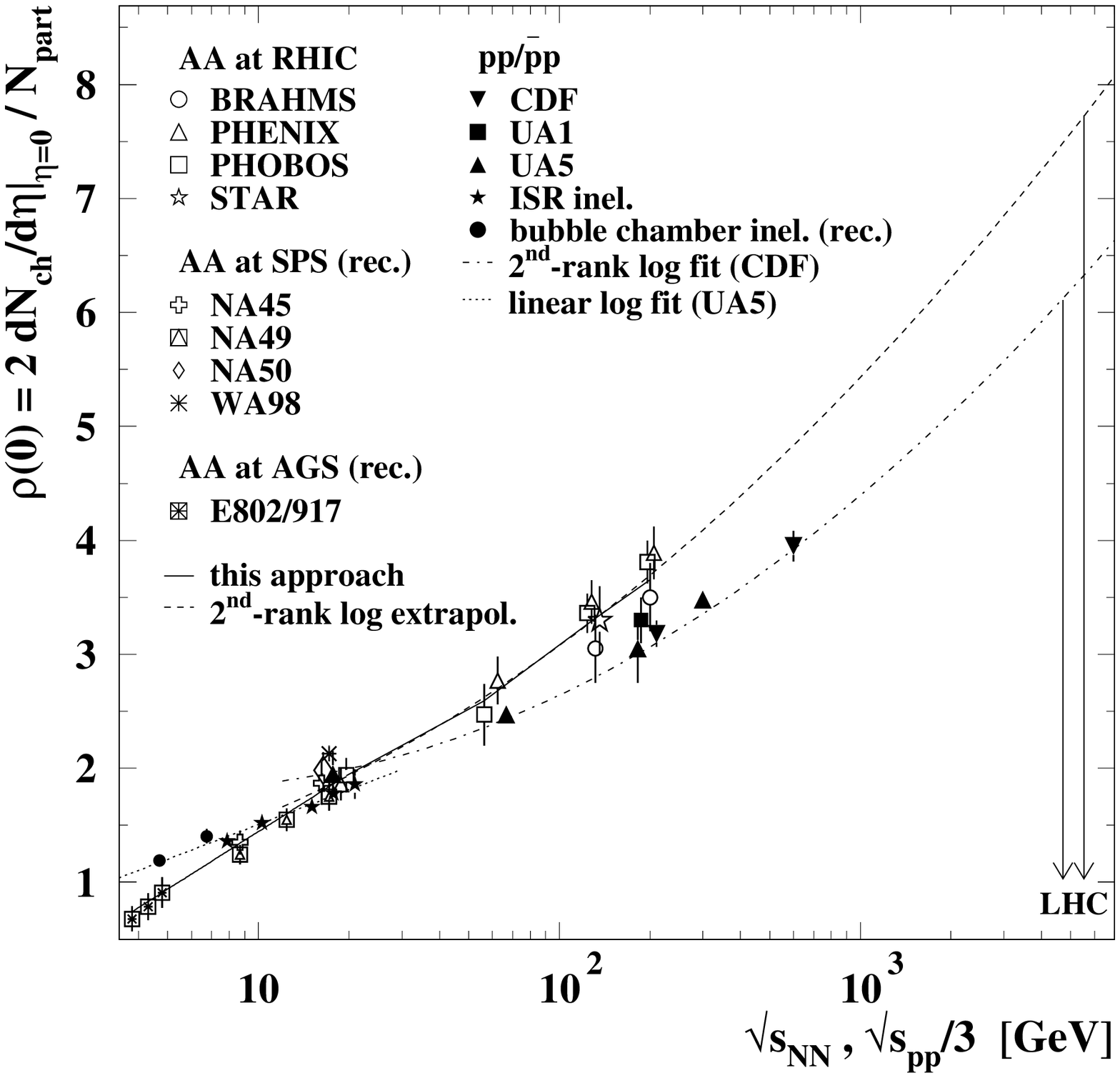}
 \vs{-.7cm}
\caption{   
Pseudorapidity density $\rho(0)$ of charged particles per participant pair 
($N_{\rm part}/2$) at 
midrapidity as a function 
of the c.m. energy of 
collision. The open and combined symbols show the pseudorapidity 
density values 
\vrs\ c.m. energy per nucleon, $\Ecmn$, measured in the headon 
AA collisions 
by BRAHMS \ct{br-rev} 
 ($ \scriptscriptstyle 
 \bigcirc$), 
PHENIX \ct{phx-rev}
($ \scriptstyle 
 \triangle$), 
PHOBOS 
\ct{ph-rev} 
 ($ \scriptstyle 
 \square$), 
and STAR 
\ct{star-200} 
 ($\star$),  
and the density values recalculated \ct{phx-rev} from the 
measurements taken by 
CERES/NA45 \ct{na45} (+), NA49 \ct{na49} ($\triangleinsquare$),  
NA50  \ct{na50}  
 ($ \scriptstyle 
 \diamondsuit$) 
WA98 \ct{wa98}
 ($\timesplus$), 
E802, and E917  \ct{ags} ($\timesplussquared$).
 The nuclear data at $\Ecmn$ around 20~GeV and the RHIC data at 
$\Ecmn=130$~GeV and 200~GeV are given spread horizontally 
for clarity. 
 The solid symbols show the pseudorapidity density values \vrs\ c.m. 
energy 
$\Ecmp/3$ as 
measured 
in non-single diffractive ${\bar {\rm p}}{\rm p}$ collisions 
 by UA1 \ct{ua1} 
 ($ \scriptstyle 
 \blacksquare$), 
UA5  \ct{ua5-53900,ua5-546} 
 ($\blacktriangle$),
CDF \ct{cdf}  
 ($\blacktriangledown$), 
and from inelastic pp data from ISR \ct{isr-thome} ($ 
 \scriptstyle 
\bigstar$), and  bubble chamber 
\ct{fnalmult,fnal-rap205} 
($\bullet$) experiments (the latter as recalculated in \ct{ua5-53900}).  
 The solid  line connects the predictions
 from Eq.~(\ref{prap0}). The dashed line 
gives the fit to the calculations using the 2nd order 
log-polynomial fit 
function analogous to that
 used \ct{cdf}
 in
 \app\ data. 
 The
 fit function from \ct{cdf} is shown by the dashed-dotted line. The dotted 
line shows the linear 
log 
 approximation
of UA5 to inelastic events \ct{ua5-53900}.   
 The arrows show the LHC expectations.
Note that \ep\ data  at $\Ecme=$~14~GeV to 200~GeV (not shown) 
follows 
 the heavy-ion data  \ct{ph-rev}.
\vs{-.7cm} 
}
\label{fig:rap0}
\end{figure}

{\bf 3.} 
In Fig.~\ref{fig:rap0}, 
we compare 
the
pseudorapidity densities per participant pair at midrapidity as  
 a function of $\Ecmn$ 
 from
headon AA collisions at RHIC, CERN SPS and AGS   
 to those of
pp/\app\ data
 from CERN 
and Fermilab 
plotted \vrs\ $\Ecmp/3$. 
Again one can see that up to the existing $\Ecmn$ the 
 data  from hadronic and 
nuclear experiments are close to each other being consistent with our
 interpretation. The measurements from the two types of collisions 
coincide at  $8<\Ecmn<20$~GeV and 
 are of the magnitude of the spread of AA data 
points
at  200~GeV. 
However, above and below the 8-20~GeV region,
there are visible differences in   
the 
midrapidity $\eta$-density values 
from AA \vrs\ pp data.  
   These indicate that,
in contrast to the
 mean multiplicity
which is a more global observable, 
the
midrapidity density depends on some additional factor. 
As the densities 
are measured in the very central $\eta$-region, where the
participants longitudinal velocities are zeroed, 
it is natural 
to assume 
that this factor is related 
to
the size  of the Lorentz-contracted volume of the initial thermalized
system determined by participants.

To take into account  the corresponding correction, let us 
consider our 
picture in the 
framework of the   
Landau model which is close to our description. 
 Then, one finds  for the ratio of the normalised charged particle
rapidity density $\rho(y)=(2/N_{\rm part})dN_{\rm ch}/dy$
at the 
midrapidity value $y=0$ 
in 
AA reaction, $\rho_{\rm NN}$, to the density $\rho_{\rm pp}$ 
in 
pp/\app\ interaction,
 \vspace*{-.34cm}
 \begin{equation}
{\rho_{\rm NN}(0)}/{\rho_{\rm pp}(0)}=
  {2\,N_{\rm ch}}
({{L_{\rm pp}}/{L_{\rm NN}}})^{1/2}
 \, 
/\big( {N_{\rm part}\, N^{\rm pp}_{\rm ch}} \big)
\,. 
\label{rap0}
 \vspace*{-.2cm}
\end{equation}
 \nopar
Here,  
$N_{\rm ch}$ ($N_{\rm ch}^{\rm pp}$) 
 is the 
multiplicity in AA (pp/\app) collision, 
$L= \ln [\sqrt {s}/(2m)]$,  and $m$ is the participant mass, \eg\ the 
proton mass $m_{\rm p}$ in AA reaction.  
 According to our interpretation, we compare in the ratio (\ref{rap0}) 
 $\rho_{\rm NN}(0)$
to   $\rho_{\rm pp}(0)$
at $\Ecmn = \Ecmp/3$ and consider a 
constituent quark of mass $\frac{1}{3}m_{\rm p}$ as a participant in 
pp/\app\ 
collisions and a proton as an effectively structureless participant in  
headon AA 
collisions.
 Then, Eq.~(\ref{rap0}) reads:
 \vspace*{-.2cm}
 \begin{equation}
  \rho_{\rm NN}(0)= 
{2\,N_{\rm ch}}\,
\rho_{\rm pp}(0) \,   
 \,
\sqrt
{1-{4 \ln 3}/{\ln\, (4 m_{\rm p}^2/s_{\rm NN})} }\,
\big/
  \big( {N_{\rm part}\, N^{\rm pp}_{\rm ch}}\big)
\,. 
\label{prap0}
 \vspace*{-.2cm}
\end{equation}

Using the fact that the transformation factor from $y$
to $\eta$ does not 
influence the above ratio and substituting the 
 multiplicity 
values 
 from  
Fig.~\ref{fig:multshe} and of $\rho_{\rm pp}(0)$
 from Fig.~\ref{fig:rap0} into Eq.~(\ref{prap0}), one obtains 
the 
values of $\rho_{\rm NN}(0)$,  
 displayed in 
Fig.~\ref{fig:rap0} 
by solid line. One can see that the 
correction made provides  good agreement between the  
calculated 
$\rho_{\rm NN}(0)$ 
values  and the
data. 
 Eq.~(\ref{prap0}) shows  the importance of the correction
for the 
participant 
 type
 to be introduced as argued above.
One can see that our calculations account also for different
types of rise of AA data below and above SPS region.
  Note  that the same two regions 
recently have 
been 
indicated by PHENIX \ct{phx-rev} from
the ratio of the midrapidity {\it transverse energy} density to the 
pseudorapidity 
density.
From these findings,  one can expect 
the midrapidity transverse energy densities in pp/\app\ and 
headon AA collisions to be similar due to the description proposed here.  
 Also, the SPS transition region properties discussed 
by  NA49 
\ct{na49-mult},  can be treated without any 
additional assumptions.
 \smallskip

{\bf 4.} 
 To estimate $\rho_{\rm NN}(0)$ for 
$\Ecmn>200$~GeV, we 
extrapolated the values of
Eq.~(\ref{prap0}) 
 utilizing 
the function found \ct{cdf} to fit well the 
\app\ 
data.  The predictions for $\rho_{\rm NN}(0)$ and the fit for 
\app\ 
data are shown in Fig.~\ref{fig:rap0} by dashed and dashed-dotted lines, 
respectively.

The obtained $\rho_{\rm NN}(0)$ 
 show faster rise with $\Ecmn$
than $\rho_{\rm pp}(0)$.
 Our calculations, 
sharing the behaviour 
at SPS--RHIC energies with that up to the LHC ones, 
give
 $\rho_{\rm NN}(0)\approx 7.7$ for
 LHC. From the CDF 
fit \ct{cdf} and assuming it covers LHC  energies, one finds 
 $\rho_{\rm pp}(0)\approx 6.1$.  
 Our $\rho_{\rm NN}(0)$ value for LHC is consistent
with
that of $\approx 6.1$ given in the PHENIX extrapolation 
\ct{phx-rev} within 
1-2 particle 
error acceptable in the calculations we made. Our result is in a good 
agreement with the best ATLAS Monte Carlo tune \ct{atlas}.
 Noticing that  $\Ecmn$ is near to $\Ecmp/3$ at LHC,  
the close values of   $\rho_{\rm NN}(0)$ and  $\rho_{\rm pp}(0)$, 
predicted for LHC  by us and estimated 
independently in \ct{phx-rev,atlas}, demonstrates 
experimentally 
 grounded description and predictive ability of our interpretation.

 Solving 
Eq.~(\ref{prap0}) for  $N_{\rm 
ch}/(0.5 N_{\rm part})$  we predict
the AA mean multiplicity energy 
dependence at $\Ecmn>200$~GeV.  In this calculations, we use the 
fits of 
$\rho_{\rm pp}(0)$ \ct{cdf} and $N^{\rm pp}_{\rm ch}$ \ct{ua5-53900} and 
our approximation for
$\rho_{\rm NN}(0)$, all  shown in 
Figs.~\ref{fig:multshe} and
\ref{fig:rap0}. From the resulted curve for 
$N_{\rm ch}/(0.5 N_{\rm part})$ 
given in 
Fig.~\ref{fig:multshe}, 
one finds that the value obtained for LHC  is just about 
$10\%$ above the $N^{\rm pp}_{\rm ch}(\Ecmp)$ fit \ct{ua5-546} prediction 
for 
LHC and about 3.3 
times 
larger the AA RHIC data at 
$\Ecmn=$~200~GeV. 
Again, this number is 
comparable with 
the estimate made by
\ct{phx-rev} and points out to no evidence for change to another regime 
  as the $\Ecmn$  increases by about two 
magnitudes from  
 the top SPS energy.      
 Nevertheless, one can see that the data obtained at the highest  RHIC 
energy  give  a hint to 
some 
border-like behaviour of the mean multiplicity where the 
pp/\app\ 
 data saturate the nuclear data, and another transition energy region  
is possible to be 
found (as at 
low energies). This  makes  
AA 
experiments 
at $\Ecmn>200$~GeV of particular interest.
\smallskip

{\bf 5.} At the end, let us dwell on the following. 

From 
our description, the mean 
multiplicity in 
{\it nucleon}-nucleus collisions  is predicted to be of the same values 
as 
that 
in pp/\app\ data, and, moreover, almost no  centrality dependence is 
expected 
for such type of interactions \ct{my}. 
These predictions are well 
confirmed by various data from
hadron-nucleus collisions at $\Ecmn
\approx$~10--20~GeV to 
recent RHIC dAu data 
at 200~GeV 
\ct{ph-rev}.
 The same seems to be correct also for the pseudorapidity density at 
midrapidity, which is already supported to be a trend \ct{ph-rev}. 
These findings remind about similar conclusions made about two decades ago 
\ct{feinberg1}.
 
 The recent observation \ct{rachid,CuCu} made at RHIC for multihadron 
data 
from CuCu 
collisions to not change compared to same c.m. energy AuAu data 
when scaled 
for the same participant numbers is also understood due to 
our description as already mentioned. Indeed, for the same number 
of 
participants, no difference in 
the
bulk variables is expected as one moves from one type of  
(identical) colliding nuclei to another one at the same c.m.
energy as soon as the same
energy is deposited into the thermalization zone.
 Note that the proper definition of participants and, thus, of the 
energy 
available 
for 
particle production, as we discuss here, allows scaling
within the
constituent 
quark picture to be applied
\ct{rachid,voloshin,ind}
 to  model the 
multihadron
data at RHIC for different observables.  
 \smallskip


One of us (EKGS) is grateful to Organizers for
invitation and partial 
financial support.








\vspace*{-2.cm}



\end{document}